\def\rfr#1{equation~(\ref{#1})}

\def\virg#1{``#1"}

\def\eqi{\begin{equation}}
\def\eqf{\end{equation}}
\def\eqia{\begin{eqnarray}}
\def\eqfa{\end{eqnarray}}

\def\rp#1#2{{#1\over#2}}
\def\lb#1{\label{#1}}

\def\bds#1{\boldsymbol{#1}}

\def\ton#1{\left(#1\right)}
\def\qua#1{\left[#1\right]}
\def\grf#1{\left\{#1\right\}}

\documentclass[onecolumn]{aastex}
\usepackage{morefloats}
\usepackage[title]{appendix}
\usepackage{hyperref}
\usepackage{booktabs}
\usepackage[table,xcdraw]{xcolor}
\usepackage{multirow}
\usepackage{rotating,tabularx}
\usepackage{float}
\usepackage{enumerate}
\usepackage{rotating}
\usepackage[polutonikogreek,english]{babel}
\usepackage{amsmath,starfont,textgreek,w-greek,wasysym}
\usepackage{amsthm}
\usepackage{amscd,lineno}
\usepackage{amssymb,dsfont}
\usepackage{graphicx,epsfig}
\usepackage{txfonts}
\usepackage{xr-hyper}

\RequirePackage{color}

\newcommand{\emaila}{lorenzo.iorio@libero.it}

\allowdisplaybreaks[1]

\begin{document}

\title{Are the planetary orbital effects of the Solar dark matter wake detectable?}

\shortauthors{L. Iorio}

\author{Lorenzo Iorio\altaffilmark{1} }
\affil{Ministero dell'Istruzione, dell'Universit\`{a} e della Ricerca
(M.I.U.R.)-Istruzione
\\ Permanent address for correspondence: Viale Unit\`{a} di Italia 68, 70125, Bari (BA),
Italy}

\email{\emaila}

\begin{abstract}
Recently, there has been some discussion in the literature about the effects of the anisotropy in the spatial density of dark matter in the Solar neighbourhood arising from the motion of the Sun through the Galactic halo. In particular, questions have been asked about the orbital motions of the solar system's planets and whether these motions can be effectively constrained by the radiotechnical observations collected by  \textit{Cassini}. I show that the semilatus rectum $p$, the eccentricity $e$, the inclination $I$, the longitude of the ascending node $\Omega$, the longitude of perihelion $\varpi$, and the mean anomaly at epoch $\eta$ of a test particle of a restricted two-body system affected by the gravity of a dark matter wake undergo secular rates of change. In the case of Saturn, they are completely negligible, being at the order of $\simeq 0.1$ millimeter per century and  $\simeq 0.05-2$ nanoarcseconds per century: the current (formal) accuracy level in constraining any anomalous orbital precessions is of the order of $\simeq 0.002-2$ milliarcseconds per century for Saturn. I also numerically simulate the Earth-Saturn range signature $\Delta\rho(t)$ due to the dark matter wake over the same  time span (2004-2017) as covered by the \textit{Cassini} data record. I find that it is as low as $\simeq 0.1-0.2\,\mathrm{m}$, while the existing range residuals, computed by astronomers without modeling any dark matter wake effect, are of the order of $\simeq 30\,\mathrm{m}$. The local dark matter density $\varrho_\mathrm{DM}$ would need to be larger than the currently accepted value of $\varrho_\mathrm{DM}=0.018\,\mathrm{M}_\odot\,\mathrm{pc}^{-3}$ by a factor of $2.5\times 10^6$ in order to induce a geocentric Kronian range signature large enough to make it discernible in the present-day residuals.
\end{abstract}

keywords{
ephemerides - celestial mechanics - space vehicles  - dark matter - gravitation - planets and satellites: dynamical evolution and stability
}
\section{Introduction}
The local density of dark matter (DM) at the Sun's location in the Galaxy may not be spherically symmetric because the Sun, in its motion through the Galactic halo, is expected to create a trailing DM wake \citep{2019arXiv190110605H,2019MNRAS.tmp.1533B}. Thus, DM would be overdense behind the Sun inducing an asymmetry which, according to some researchers \citep{2019arXiv190110605H,2019MNRAS.tmp.1533B}, may allow for tighter constraints on the DM density due to its effects on the orbital motions of the planets of our solar system.

The perturbing gravitational acceleration experienced by a test particle orbiting our star which moves in the DM background can approximately be written in some coordinate system as \citep{1983A&A...117....9M, 2019arXiv190110605H,2019MNRAS.tmp.1533B}
\eqi
{\bds A}_\mathrm{DM} = -\rp{4\,\uppi\,G^2\,\varrho_\mathrm{DM}\,\mathrm{M}_\odot}{\sigma^2}\,\qua{0.21\,\ln\ton{\rp{r\,\mathrm{v}^2_\odot}{2\mu_\odot}} + 0.44\,\rp{{\bds{\hat{v}}_\odot}\bds\cdot{\bds{\hat{r}}} }{\left| {\bds{\hat{v}}_\odot}\bds\cdot{\bds{\hat{r}}} \right|}}\bds{\hat{v}}_\odot.\lb{wakeacc}
\eqf
In \rfr{wakeacc}, $G$ is the Newtonian gravitational constant, $\mu_\odot\doteq G\,\mathrm{M}_\odot$ is the Sun's gravitational parameter, $\mathrm{M}_\odot$ is its mass, ${\bds v}_\odot$ is the velocity of the Sun's motion through the Galactic DM halo, $\mathrm{v}_\odot = \left|{\bds v}_\odot\right|$ is its speed, $\varrho_\mathrm{DM}$ is the unperturbed local DM density, $\sigma = \mathrm{v}_\odot/\sqrt{2}$ is its one-dimensional velocity dispersion, and $\bds{\hat{r}}$ is the versor of the heliocentric position vector $\bds r$ of the planet.
The planetary observations are processed in the International Celestial Reference System (ICRS), whose fundamental plane is the celestial equator at the reference epoch J2000. Thus, ${\bds{\hat{v}}}_\odot$ must be transformed from the Galactic coordinate system (GalCS), which is a right-handed one whose $x$ axis points towards the Galactic Center, the $z$ axis is directed towards the North Galactic Pole (NGP), and the $y$ axis is aligned with the local direction of the large scale ordered rotation of the Galactic disk, to the equatorial system of ICRS. To the accuracy level required by the problem at hand, such a task can straightforwardly be accomplished with the inverse of the matrix $\mathcal{N}$ in \citet{2011A&A...536A.102L}. In the GalCS, ${\bds v}_\odot$ is
\eqi
{\bds v}_\odot^\mathrm{GalCS}= \grf{U_\odot,\,V_\odot+\Theta_\odot,\,W_\odot},
\eqf
where \citep{2018RNAAS...2c.156M} $\Theta_\odot = 233.3\,\mathrm{km\,s}^{-1}$ is the circular speed of the Local Standard of Rest (LSR), and \citep{2010MNRAS.403.1829S}
$U_\odot =11.1 \,\mathrm{km\,s}^{-1},\,V_\odot = 12.24\,\mathrm{km\,s}^{-1},\,W_\odot = 7.25\,\mathrm{km\,s}^{-1}$ are the components of the velocity of the Sun with respect to the LSR itself. Thus, the unit vector of the Sun's Galactic velocity, referred to the ICRS, turns out to be
\eqi
{\bds{\hat{v}}}_\odot = \grf{0.45574,\,-0.494244,\,0.740287}.
\eqf

In this Letter, I investigate in detail the orbital effects of \rfr{wakeacc} on the planets of our solar system without any a priori simplifying assumptions on their orbital configuration in order to make a consistent and unambiguous comparison with the observable quantities actually delivered by the astronomers. For other investigations about Saturn, performed with different methodologies, see \citet{2019arXiv190110605H,2019MNRAS.tmp.1533B}. In particular, I numerically calculate the long-term rates of change of all the Keplerian orbital elements of a test particle orbiting a primary under the influence of the perturbing acceleration of \rfr{wakeacc}. I apply my results to Saturn, and compare its DM-induced secular rates with the most recent bounds on any anomalous extra-precessions of it existing in the literature.  Moreover, I numerically simulate the Earth-Saturn range signature due to \rfr{wakeacc}, and compare it with the currently available range residuals computed by the astronomers with the data collected by the \textit{Cassini} spacecraft from 2004 to 2017.
\section{The Keplerian orbital elements and the Earth-Saturn range}
Here, I investigate some of the consequences of \rfr{wakeacc} in terms of the orbital effects induced by it on the motion of a test a particle around its primary in a restricted two-body system.

\begin{figure}[htb]
\begin{center}
\centerline{
\vbox{
\begin{tabular}{cc}
\epsfysize= 5.2 cm\epsfbox{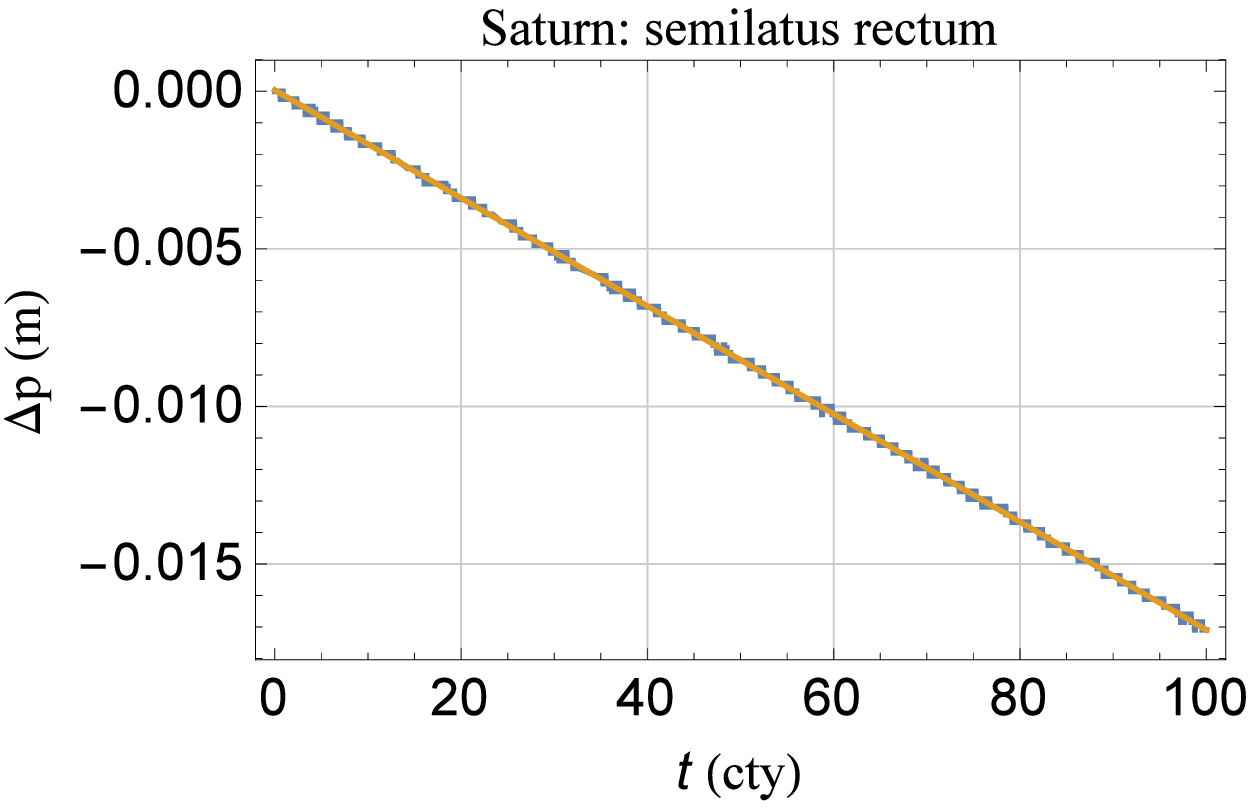} & \epsfysize= 5.2 cm\epsfbox{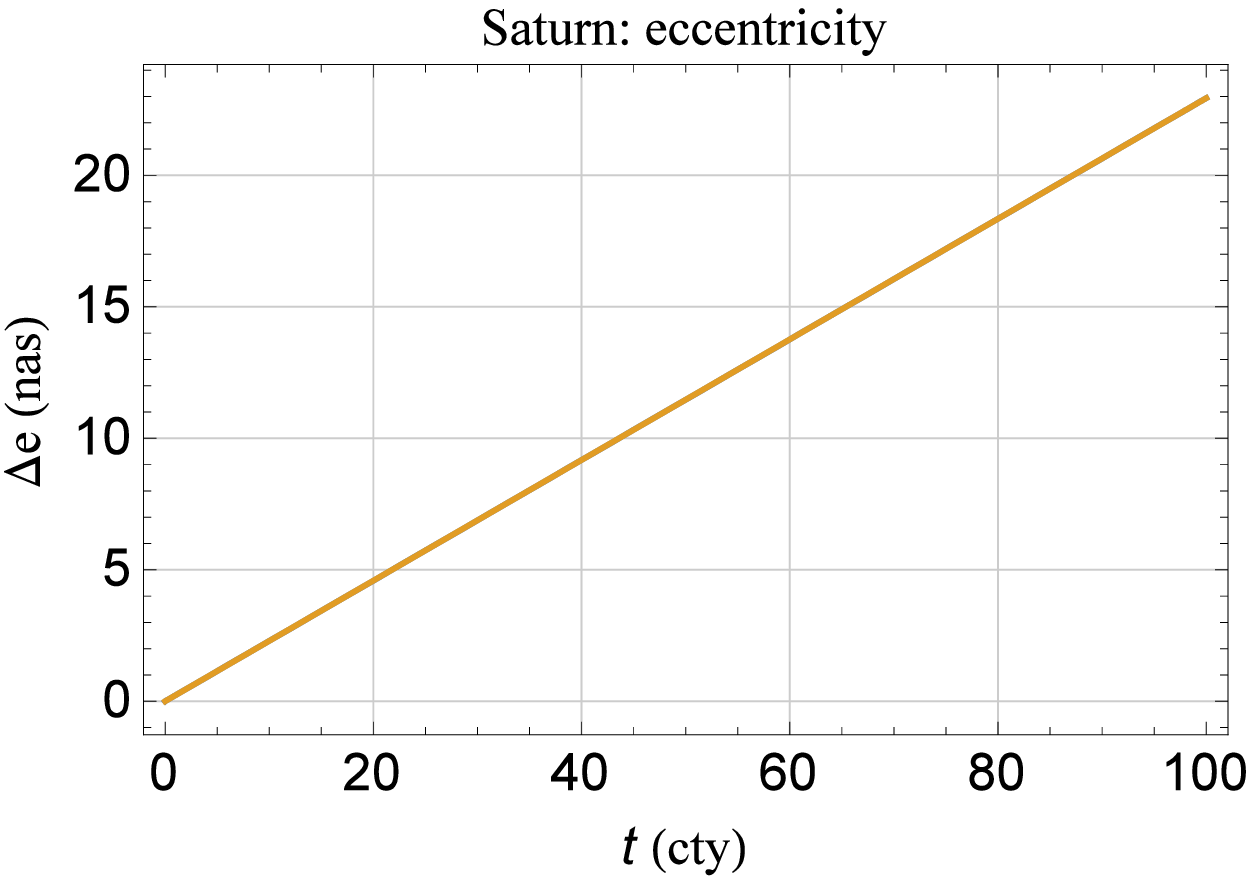}\\
\epsfysize= 5.2 cm\epsfbox{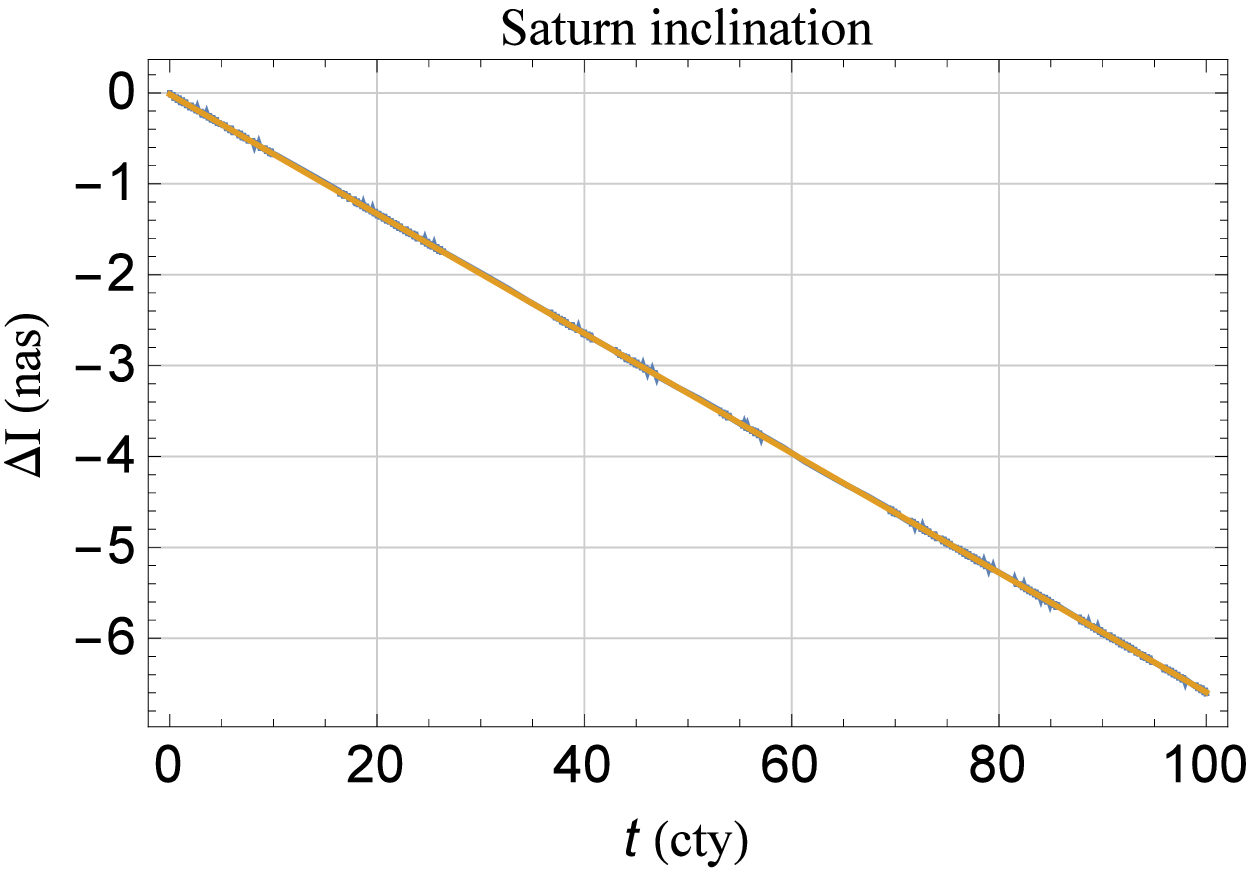} & \epsfysize= 5.2 cm\epsfbox{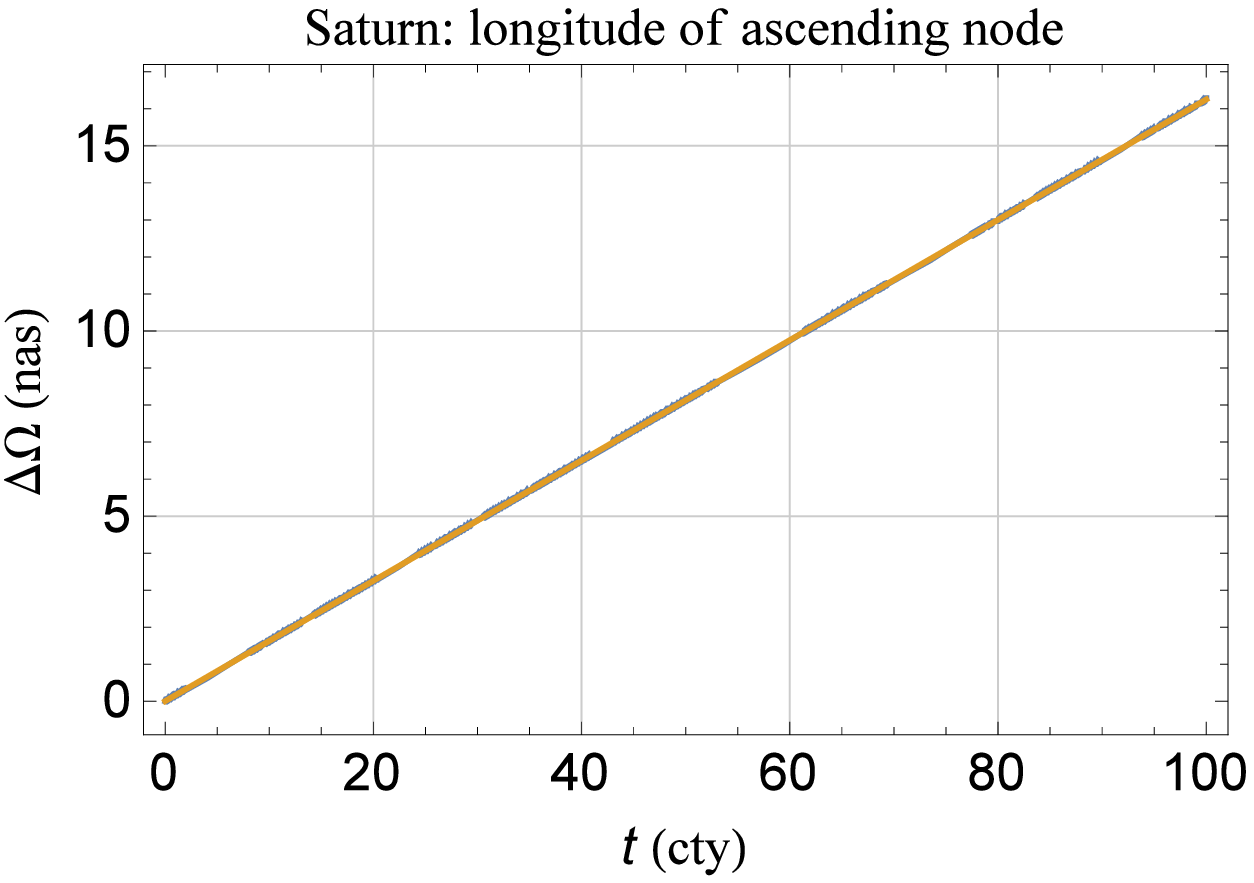}\\
\epsfysize= 5.2 cm\epsfbox{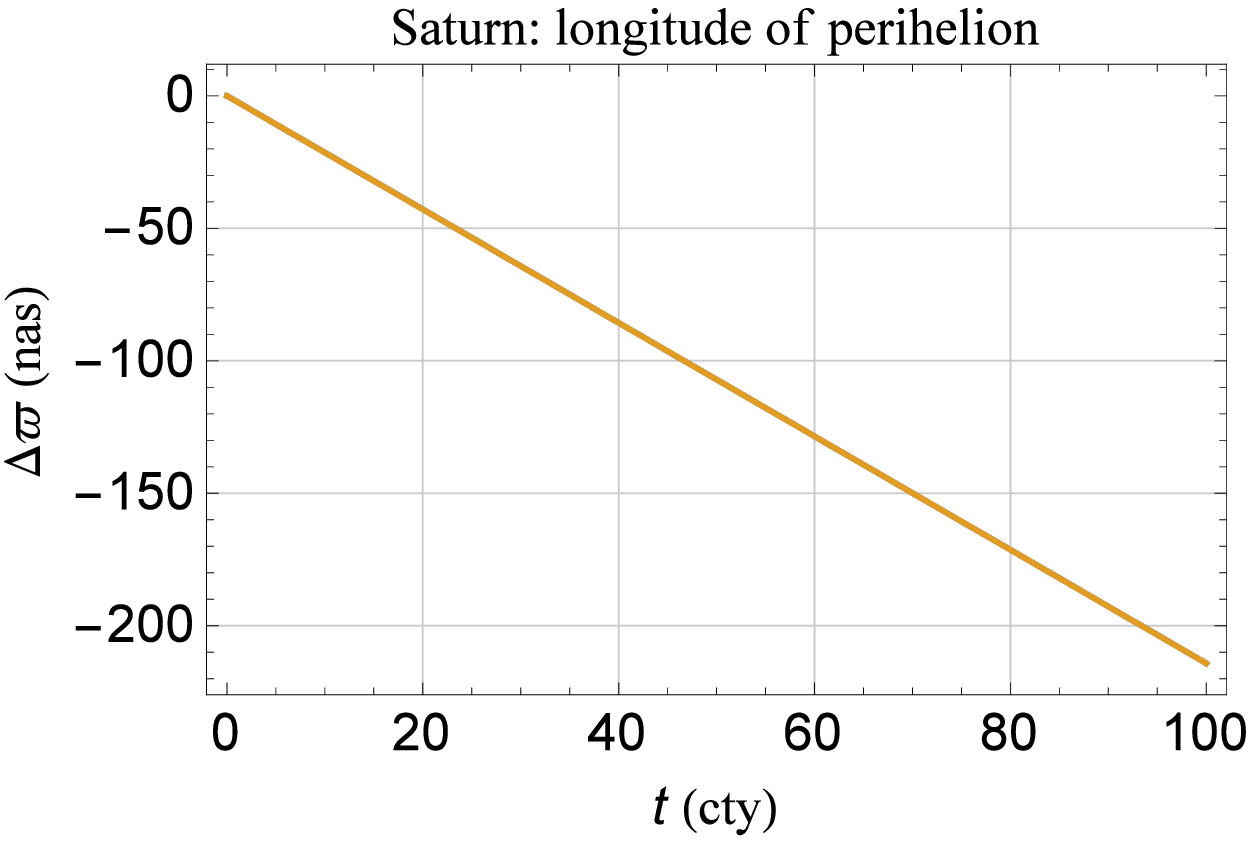} & \epsfysize= 5.2 cm\epsfbox{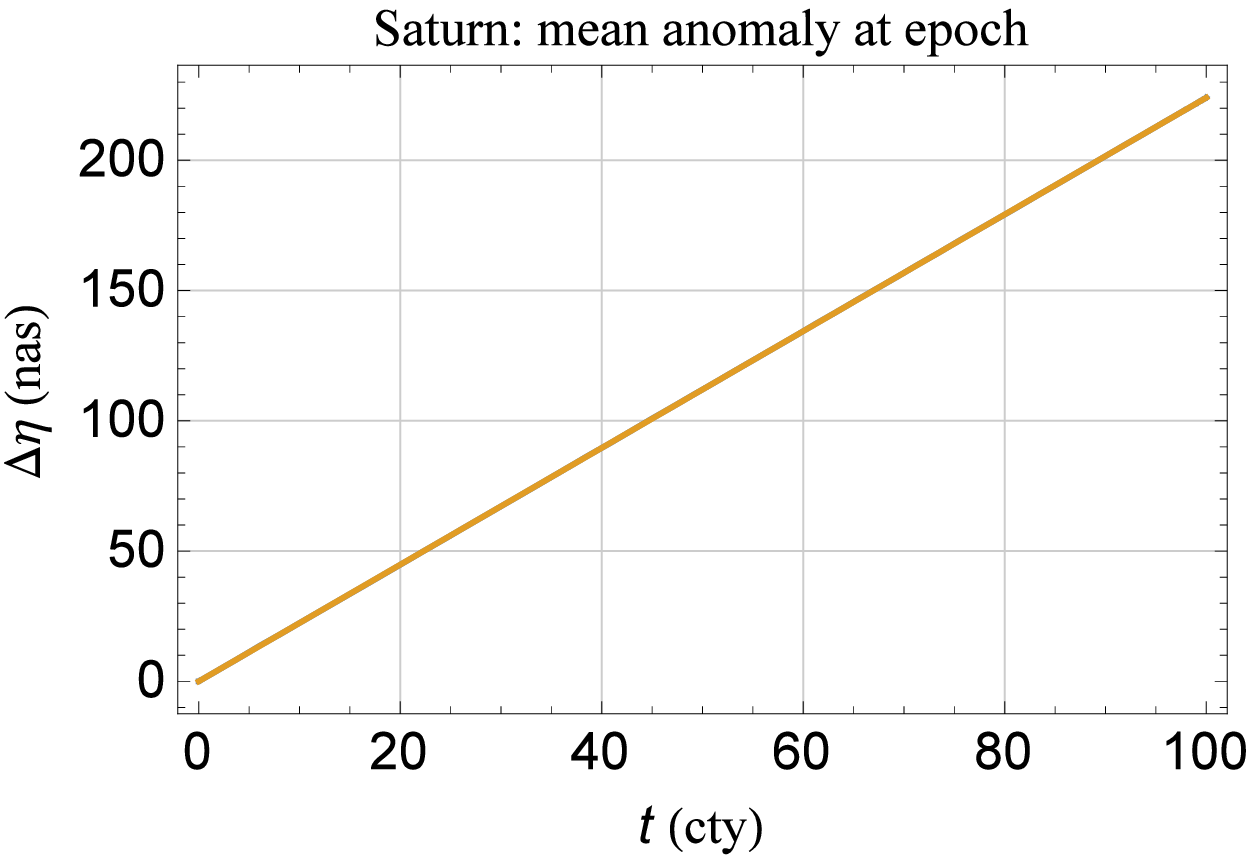}\\
\end{tabular}
}
}
\caption{
Numerically integrated shifts of the semilatus rectum $p$, eccentricity $e$, inclination $I$, longitude of the ascending node $\Omega$, longitude of perihelion $\varpi$, and mean anomaly at epoch $\eta$ of Saturn induced by the Solar DM wake acceleration of \rfr{wakeacc} over a time span of 100 centuries. The units are metres for $p$ and nanoarcseconds (nas) for all the other orbital elements. They were  obtained for each orbital element as differences between two time series calculated by numerically integrating the barycentric Kronian equations of motion in Cartesian rectangular coordinates  with and without \rfr{wakeacc} for $\varrho_\mathrm{DM} =  0.018\,\mathrm{M}_\odot\,\mathrm{pc}^{-3}$ \citep{2019MNRAS.tmp.1533B}. The initial conditions, referred to the celestial equator at the reference epoch J2000, were retrieved from the WEB interface HORIZONS by NASA JPL; they were the same for both the integrations. The Sun's Galactic velocity ${\bds v}_\odot$ was transformed to the International Celestial Reference System (ICRS). The slopes of the resulting secular trends are listed in Table\,\ref{tavola1}. }\label{figura1}
\end{center}
\end{figure}

In principle, it would be possible to analytically work out the long-term rates of change of its Keplerian orbital elements with the Gauss perturbative equations applied to \rfr{wakeacc} by averaging their right-hand-sides, evaluated onto an unperturbed Keplerian ellipse as reference trajectory, over an orbital period. In view of how cumbersome such an approach is, however I will take a numerical approach. In particular, I simultaneously integrate the equations of motion of, say, Saturn in Cartesian rectangular coordinates and the Gauss equations for each orbital element with and without \rfr{wakeacc} over a time span as long as 100 centuries in order to clearly identify the sought features of motion: both runs share the same initial conditions, as retrieved from the Internet from the WEB interface HORIZONS maintained by NASA Jet Propulsion Laboratory (JPL).
For consistency reasons with the planetary data reductions available in the literature, I use the equatorial coordinates of the ICRS. Then, for each orbital element, Fig.\,\ref{figura1} plots the time series resulting from the difference between the runs with and without \rfr{wakeacc}. Finally, I fit a linear model to its numerically produced signal, and estimate its slope: the results are given in Table\,\ref{tavola1}.
\clearpage{}
\begin{table}
\caption{Estimated slopes of the secular trends induced by the Solar DM wake acceleration of \rfr{wakeacc}
for $\varrho_\mathrm{DM} =  0.018\,\mathrm{M}_\odot\,\mathrm{pc}^{-3}$ \citep{2019MNRAS.tmp.1533B} on the semilatus rectum $p$, eccentricity $e$, inclination $I$, longitude of the ascending node $\Omega$, longitude of perihelion $\varpi$, and mean anomaly at epoch $\eta$ of Saturn according to Fig.\,\ref{figura1}. The units are millimetres per century $\ton{\mathrm{mm\,cty}^{-1}}$ for $p$, and nanoarcseconds per century $\ton{\mathrm{nas\,cty}^{-1}}$ for all the other orbital elements.
}\lb{tavola1}
\begin{center}
\small{
\begin{tabular}{|l|l|l|l|l|l|}
\hline
    $\dot p\,\left(\textrm{mm}\,\textrm{cty}^{-1}\right)$
  & $\dot e\,\left(\textrm{nas}\,\textrm{cty}^{-1}\right)$
  & $\dot I\,\left(\textrm{nas}\,\textrm{cty}^{-1}\right)$
  & $\dot\Omega\,\left(\textrm{nas}\,\textrm{cty}^{-1}\right)$
  & $\dot\varpi\,\left(\textrm{nas}\,\textrm{cty}^{-1}\right)$
  & $\dot\eta\,\left(\textrm{nas}\,\textrm{cty}^{-1}\right)$ \\
\hline
 $-0.1$ & $0.2$ & $-0.06$ & $0.2$ & $-2.1$ & $2.2$\\
\hline
\end{tabular}
}
\end{center}
\end{table}
It turns out that the impact of the Sun's DM wake on Saturn's motion is totally negligible. Indeed, its predicted orbital effects are as low as $\simeq 0.1\,\mathrm{millimeters\,per\,century}\,\ton{\mathrm{mm\,cty}^{-1}}$ and
$\simeq 0.05-2\,\mathrm{\mathrm{nanoarcseconds\,per\,century}}\,\ton{\mathrm{nas\,cty}^{-1}}$. On the other hand, the present-day formal accuracies in constraining any anomalous orbital rate of change of Saturn amount to $\simeq 17\,\mathrm{m\,cty}^{-1}$ and $\simeq 0.002-2\,\mathrm{milliarcseconds\,per\,century}\,\ton{\mathrm{mas\,cty}^{-1}}$, respectively, as tentatively calculated by \citet{2019AJ....157..220I} on the basis of the latest results by \citet{2018AstL...44..554P} with the recent EPM2017 ephemerides.

\begin{figure}[htb]
\begin{center}
\centerline{
\vbox{
\begin{tabular}{c}
\epsfysize= 7.0 cm\epsfbox{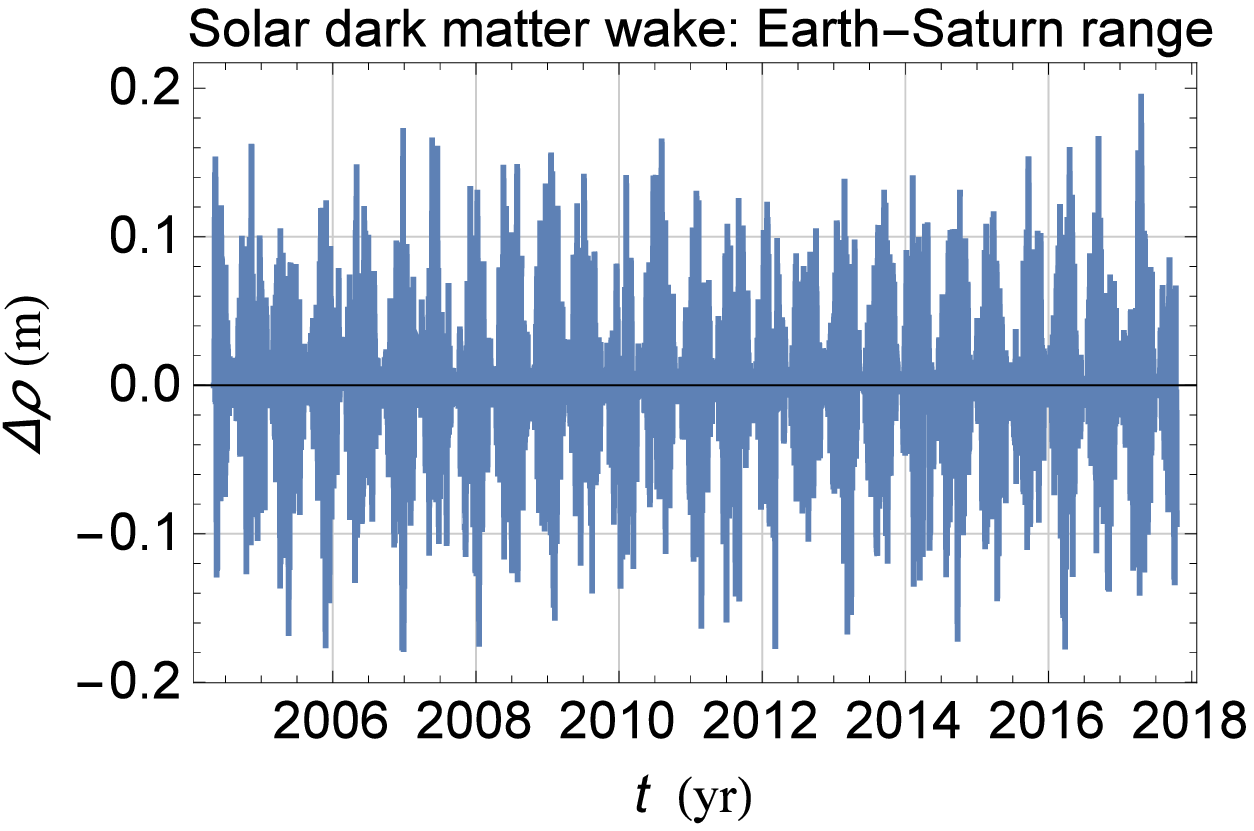}\\
\epsfysize= 7.0 cm\epsfbox{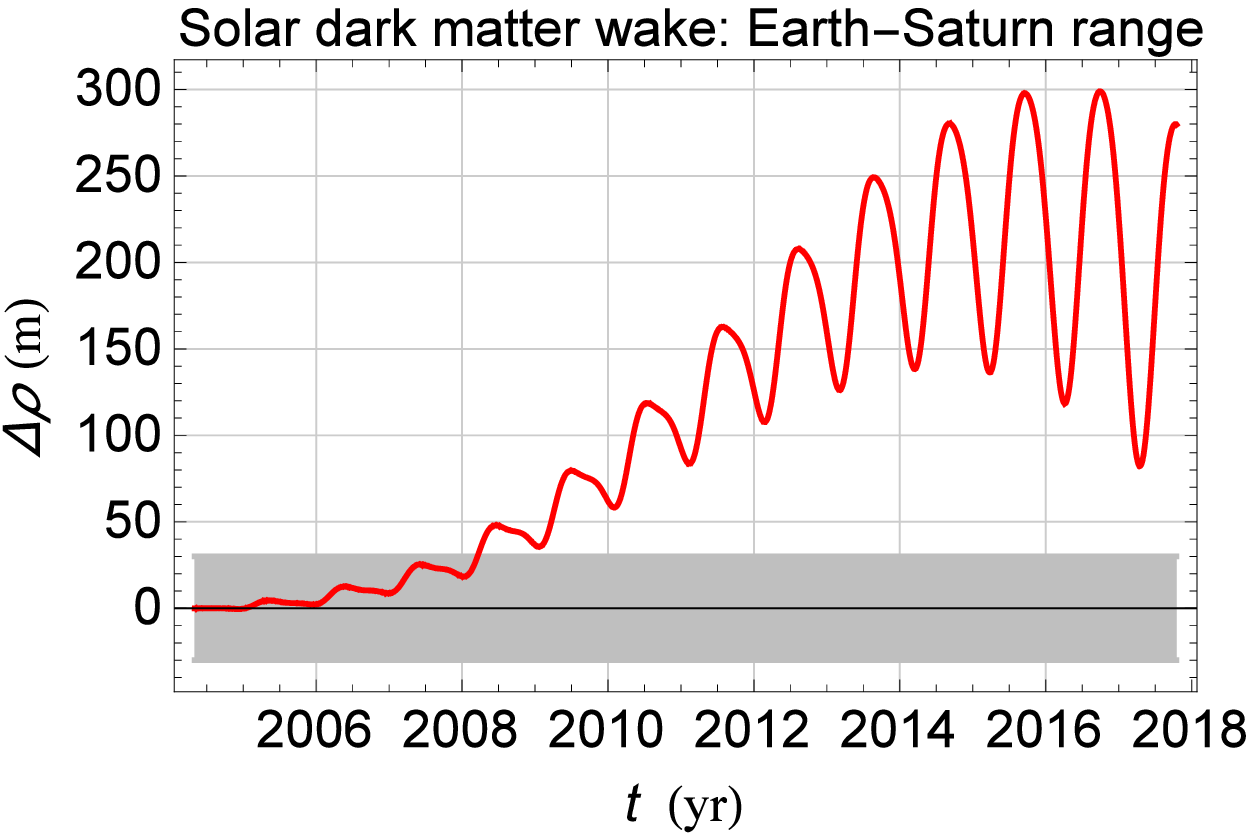}\\
\end{tabular}
}
}
\caption{
Numerically simulated Earth-Saturn range signature $\Delta\rho(t)$ induced by the Solar DM wake acceleration of \rfr{wakeacc} over a time span 13 yr long covering the time spent by the \textit{Cassini} spacecraft in the Kronian system. It was  obtained as a difference between two time series of
$\rho(t)=\sqrt{\ton{x_\mathrm{Sat}(t)-x_\oplus(t)}^2 + \ton{y_\mathrm{Sat}(t)-y _\oplus(t)}^2 + \ton{z_\mathrm{Sat}(t)- z_\oplus(t)}^2}$
calculated by numerically integrating the barycentric equations of motion in Cartesian rectangular coordinates of all the major bodies of the solar system from 2004 April 1 to 2017 September 15 with and without \rfr{wakeacc} for $\varrho_\mathrm{DM} = \varrho_\mathrm{DM}^0= 0.018\,\mathrm{M}_\odot\,\mathrm{pc}^{-3}$ \citep{2019MNRAS.tmp.1533B} (upper panel) and $\varrho_\mathrm{DM}= 2.5\times 10^6\,\varrho^0_\mathrm{DM}$ (lower panel). The initial conditions, corresponding to 2004 April 1 and referred to the celestial equator at the reference epoch J2000, were retrieved from the WEB interface HORIZONS by NASA JPL; they were the same for both the integrations which share also the entire standard $N$-body dynamical models to the first post-Newtonian level. The Sun's Galactic velocity ${\bds v}_\odot$ was transformed to the International Celestial Reference System (ICRS). The gray shaded horizontal band in the lower panel has a semi-amplitude of $30\,\mathrm{m}$, and represents the \virg{standard} post-fit range residuals of Saturn produced by processing the \textit{Cassini} data without explicitly modeling \rfr{wakeacc} \citep{2017NSTIM.108.....V}. }\label{figura2}
\end{center}
\end{figure}

I also looked at the geocentric Kronian range by numerically producing a simulated time series $\Delta\rho(t)$ caused by \rfr{wakeacc} over the same time span (2004-2017) of the data record collected by the \textit{Cassini} spacecraft during its long-lasting tour in the system of the ringed planet. The time series was obtained from a simultaneous numerical integration of the barycentric equations of motion of all the major bodies of the solar system from 2004 April 1 to 2017 September 15. Two runs, sharing the same initial conditions and standard dynamical models accurate to the first post-Newtonian level, with the exception of \rfr{wakeacc} which was turned off in one of them, were performed. Then, two time series for the Earth-Saturn range were calculated, and their difference was taken as representative of $\Delta\rho(t)$ and plotted in Fig.\,\ref{figura2}.
\clearpage{}
From its upper panel, it can be noted that the expected DM-induced effect on the Earth-Saturn range is as little as $\simeq 0.1-0.2\,\mathrm{m}$; the range residuals currently available, computed by the astronomers without explicitly modeling \rfr{wakeacc}, are as large as $\simeq 30\,\mathrm{m}$ \citep{2017NSTIM.108.....V}. The lower panel of Fig.\,\ref{figura2} shows that, in order to have an anomalous signal sufficiently large to be, perhaps, detectable even with such non-dedicated residuals\footnote{The signature of any unmodeled effect $\mathcal{E}$, even if present in Nature, may be partially or totally removed in the data reduction procedure generating, among other things, the post-fit residuals since it may be partially absorbed in the estimation of other parameters like, e.g., the planetary masses and state vectors. This is especially true if its putative magnitude is not sufficiently greater than the measurements' accuracy. Thus, caution is in order when straightforward comparisons between a theoretically expected anomalous effect $\mathcal{E}$ and the residuals produced in non-dedicated analyses are made. Indeed, the absence of $\mathcal{E}$ in the residuals does not necessarily imply that $\mathcal{E}$ does not exist.}, the local DM density $\varrho_\mathrm{DM}$ would need to be about $2.5\times 10^6$ times larger than the currently accepted value $\varrho^0_\mathrm{DM} = 0.018\,\mathrm{M}_\odot\,\mathrm{pc}^{-3}$ \citep{2019MNRAS.tmp.1533B}; note also that, according to some estimates \citep{2018RNAAS...2c.156M}, $\varrho_\mathrm{DM}$ could even be smaller, possibly at the level of $\varrho_\mathrm{DM}\simeq 0.006\,\mathrm{M}_\odot\,\mathrm{pc}^{-3}$ level.
\section{Summary and conclusions}
The Solar DM wake acceleration of \rfr{wakeacc} induces long-term, secular rates of change on all the Keplerian orbital elements of the planets of our solar system. I numerically calculated them by integrating their equations of motion and using the Gauss perturbative equations after having rotated the velocity ${\bds v}_\odot$  of the Sun's Galactic travel from the GalCS to the ICRS, which is the coordinate system routinely used by the astronomers to process the planetary observations. For the presently accepted values of the parameters entering \rfr{wakeacc}, including the local DM density in the Sun's neighbourhood, the expected DM-induced orbital precessions of Saturn turn out to be as low as $\lesssim \mathrm{nas\,cty}^{-1}$, while the current formal uncertainties in the estimated Kronian orbital rates are at the $\simeq 0.002-2\,\mathrm{mas\,cty}^{-1}$ level.

I also simulated the Earth-Saturn range signature due to \rfr{wakeacc} over the same time span as covered by the data collected by the \textit{Cassini} spacecraft (2004-2017) whose residuals, computed by the astronomers without modeling any DM perturbations, are as large as $\simeq 30\,\mathrm{m}$. My numerically produced range time series, calculated with the values of the parameters of \rfr{wakeacc} found in the literature, is as low as $\simeq 0.1-0.2\,\mathrm{m}$. I demonstrated that the local DM density should be about a million times greater than its currently accepted value to create a range signal so large that it could not have escaped measurement even with the conventionally produced residuals today available.

In conclusion, the expected effects of the Solar DM wake on planets are far too small to be detected, or even effectively constrained, with the current accuracy of planetary observations, being their existence quite compatible with them.
\bibliography{MS_binary_pulsar_bib,Gclockbib,semimabib,PXbib}{}

\end{document}